\documentclass[twocolumn, showpacs, nofootinbib, aps,
superscriptaddress,
prd]{revtex4}
\usepackage{graphicx,dcolumn,bm,amsfonts,amsmath,color,xcolor,amsthm}
\usepackage[colorlinks=true, citecolor=blue, linkcolor=blue,urlcolor=blue]{hyperref}
\usepackage{slashed}
\usepackage{ulem}
\usepackage{amsmath}
\usepackage{booktabs}
\usepackage{xcolor}
\usepackage{hhline}
\usepackage{multirow}

\newcommand\add[1]{{\color{black}#1}}

\allowdisplaybreaks

\begin{document}

\title{Probing the equivalence of chiral LCSRs in $D \to \pi e \nu_e$ decays and extraction of $|V_{cd}|$}
\author{Xiu-Fen Wang}
\address{Department of Physics, Guizhou Minzu University, Guiyang 550025, P.R.China}
\author{Hai-Jiang Tian}
\address{Department of Physics, Chongqing Key Laboratory for Strongly Coupled Physics, Chongqing University, Chongqing 401331, P.R. China}
\author{Yin-Long Yang}
\address{Department of Physics, Guizhou Minzu University, Guiyang 550025, P.R.China}
\author{Long Zeng}
\address{Department of Physics, Guizhou Minzu University, Guiyang 550025, P.R.China}
\address{Department of Physics, Chongqing Key Laboratory for Strongly Coupled Physics, Chongqing University, Chongqing 401331, P.R. China}
\author{Hai-Bing Fu}	
\email{fuhb@gzmu.edu.cn}
\address{Department of Physics, Guizhou Minzu University, Guiyang 550025, P.R.China}

\begin{abstract}
In the paper, we have carried out research on the $D\to\pi$ decay process. We employ two different currents to study the $D\to\pi$ transition form factors (TFFs) by using the light-cone sum rule within the framework of chiral current approach. Firstly, we follow the right-handed and left-handed currents for the correlators to present the expression of the vector form factors upto next-leading-order and leading-order accuracy, respectively. Here the twist-2 and twist-3 light-cone distribution amplitudes are constructed by the light-cone harmonic oscillator model. After exploring the TFFs into the whole physical $q^2$-region with the simplified $z$-series expansion, we obtain the branching fractions $\mathcal{B}(D^0\to \pi^-e^+\nu_e)_{\text{I}} = 0.31_{-0.05}^{+0.05}$, $\mathcal{B}(D^+\to \pi^0e^+\nu_e)_{\text{I}} = 0.39_{-0.06}^{+0.06}$, $\mathcal{B}(D^0\to \pi^-e^+\nu_e)_{\text{II}} = 0.27_{-0.03}^{+0.05}$, $\mathcal{B}(D^+\to \pi^0e^+\nu_e)_{\text{II}} = 0.34_{-0.04}^{+0.06}$, and extract the CKM matrix element $|V_{cd}|_{\text{I}} = ( 0.21^{+0.02}_{-0.02} )\times 10^{-2}$ as well as $|V_{cd}|_{\text{II}} = ( 0.23^{+0.02}_{-0.02}) \times 10^{-2}$. To verify the credibility of our calculations, these results are further compared with existing findings in the literature, showing good agreement within uncertainties.
\end{abstract}
\maketitle

\section{Introduction}
The study of the semileptonic decays of heavy mesons provides a powerful probe for simultaneously exploring the non-perturbative regions of electroweak interactions and QCD. Among these heavy mesons, the lightest meson containing the charm quark, the $D$-meson has an abundance of decay channels. Among them, the heavy-to-light decay contains a lot of information for the dynamics of weak and strong interaction, which may provide a platform to test the Standard Model(SM) more accurately. For instance, the semileptonic decay process $D \to \pi e\nu_e$ is widely used to extract the Cabibbo-Kobayashi-Maskawa (CKM) matrix element $|V_{cd}|$, a critical component of the SM. Meanwhile, in the SM of particle physics, the mixing between the quark flavors in the weak interaction is parameterized by the CKM matrix element. This application stems from the fact that the branching fractions of such associated decays are proportional to the square of the CKM matrix element $|V_{cd}|$. Moreover, the decay of heavy mesons into a light pseudoscalar meson, an electron, and its antineutrino occurs at the electroweak tree level; this characteristic makes the process far less sensitive to potential new physics effects. As a result, the $D \to \pi e\nu_e$ emerges as preferred channels for precisely probing the value of $|V_{cd}|$. Consequently, this process has been widely studied both by experimentalists and theorists.

On the experimental side, the MARKIII Collaboration made an measurement of the branching fraction for $D^0 \to \pi^- e^+ \nu_e$ channel with the value $0.39_{-0.11}^{+0.23} \pm 0.04\%$ by analysing the data taken at the near $\bar{D}D$ threshold region early in 1989~\cite{MARK-III:1989dea}. Subsequently, the BES Collaboration~\cite{BES:2004rav} reported the branching fractions for the decays $ D^0 \to \pi^- e^+ \nu_e $ determined using $ 7584 \pm 198 \pm 341 $ singly tagged $ \bar{D}^0 $ sample from the data collected around 3.773 GeV with the BES-II detector at the BEPC to yield the result of the branching fraction $\mathcal{B}\left(D^0 \to \pi^- e^+ \nu_e\right) = (0.33 \pm 0.13 \pm 0.03)\%$. Then the Belle Collaboration~\cite{Belle:2006idb} study $D^0$ decays to $\pi^-\ell^+\nu$ final states using a $282\ \text{fb}^{-1}$ data sample collected by the Belle experiment at the KEKB $e^+e^-$ collider, and they measure the absolute branching fractions to be $\mathcal{B}\left(D^0 \to \pi^- e^+ \nu_e\right) = (0.255 \pm 0.019_{\rm stat} \pm 0.016_{\rm syst})\% $ and the semi-leptonic form factors (within the modified pole model) $ f_+^{D\pi}(0) = 0.624 \pm 0.020_{\rm stat} \pm 0.030_{\rm syst} $. In 2009, CLEO Collaboration~\cite{CLEO:2009svp} present a study of the decays $ D^0 \to \pi^- e^+ \nu_e $ and $ D^+ \to \pi^0 e^+ \nu_e $ using the entire CLEO-c $ \psi(3770) \to D\bar{D} $ event sample, corresponding to an integrated luminosity of $ 818\ \text{pb}^{-1} $ and approximately 5.4 million $ D\bar{D} $ events, and report the branching fractions $ \mathcal{B}\left(D^0 \to \pi^- e^+ \nu_e\right) = (0.288 \pm 0.008 \pm 0.003)\% $ and $ \mathcal{B}\left(D^+ \to \pi^0 e^+ \nu_e\right) = (0.405 \pm 0.016 \pm 0.009)\% $. Recently, in 2015, the BESIII Collaboration~\cite{BESIII:2018nzb} carried out relevant research using a data sample corresponding to an integrated luminosity of $ 2.9\ \text{fb}^{-1} $ taken at a center-of-mass energy of $ 3.773\ \text{GeV} $ with the BESIII detector operated at the BEPCII collider. The branching fractions of $ D^0 \to \pi^- \mu^+ \nu_\mu $ and $ D^+ \to \pi^0 \mu^+ \nu_\mu $ are measured to be $ (0.272 \pm 0.008_{\rm stat} \pm 0.006_{\rm syst})\% $ and $ (0.350 \pm 0.011_{\rm stat} \pm 0.010_{\rm syst})\% $, respectively. Using these results along with previous BESIII measurements~\cite{BESIII:2015tql} of $ D^{0(+)} \to \pi^{-(0)} e^+ \nu_e $, they calculate the branching fraction ratios to be $ \mathcal{R}^0 \equiv \mathcal{B}_{D^0 \to \pi^- \mu^+ \nu_\mu} / \mathcal{B}_{D^0 \to \pi^- e^+ \nu_e} = 0.922 \pm 0.030_{\rm stat} \pm 0.022_{\rm syst} $ and $ \mathcal{R}^+ \equiv \mathcal{B}_{D^+ \to \pi^0 \mu^+ \nu_\mu} / \mathcal{B}_{D^+ \to \pi^0 e^+ \nu_e} = 0.964 \pm 0.037_{\rm stat} \pm 0.026_{\rm syst} $, which are compatible with the theoretical expectation of lepton flavor universality within $ 1.7\sigma $ and $ 0.5\sigma $, respectively. Up to now, the $D \to \pi e\nu_e$ process has been more precisely measured by a larger number of experimental groups, such as CLEO~\cite{CLEO:2007ntr}, Babar~\cite{BaBar:2014xzf} and BESIII~\cite{BESIII:2017ylw, BESIII:2018xre} Collaborations, etc. With the continuous improvement in experimental measurement precision, making accurate theoretical predictions for this decay process has become increasingly important. This holds key significance for testing lepton flavor universality at higher precision and extracting fundamental parameters such as the CKM matrix elements.

Theoretically, exclusive decay processes involve multiple hadronic matrix elements, which must be calculated using physical meson states instead of free quarks. To facilitate such calculations, a complete set of Lorentz-invariant transition form factors (TFFs) can be introduced to parametrize these QCD dynamics. Notably, the TFFs for the $D \to \pi$ transition have been computed under various approaches, such as the constituent quark model (CQM)~\cite{Melikhov:2000yu}, QCD light-cone sum rule (LCSR)~\cite{Wu:2006rd, Wang:2002zba, Khodjamirian:2000ds, Li:2012gr, Khodjamirian:2009ys}, lattice QCD (LQCD)~\cite{Lubicz:2017syv, Na:2011mc, FermilabLattice:2004ncd, FermilabLattice:2022gku, Al-Haydari:2009kal}, covariant confined quark model (CCQM)~\cite{Soni:2018adu}, relativistic quark model (RQM)~\cite{Faustov:2019mqr} etc. Among these, LCSRs, developed from the QCD sum rule technique \cite{Balitsky:1989ry}, have emerged as a powerful framework for making predictions for heavy-to-light transitions. Complementary to LQCD simulations, this approach is successfully applied to study $D$-meson decays: whereas the former is available for the high $q^2$-region, LCSR calculation is applicable for the low and intermediate $q^2$-region~\cite{Ball:2004ye, Ball:2006yd}. So in the paper, one takes LCSRs to recalculate the $D \to \pi$ TFFs. Compare with transition sum rule, the LCSR is based on the operator product expansion (OPE) near the light-cone $x^2 \approx 0$ instead of at the short distance $x \approx 0$, where the nonperturbative dynamics are parametrized by leading-twist light-cone distribution amplitude (LCDA) of increasing twists~\cite{Huang:2001xb}. Therefore, precise LCSR predictions are dependent on meson LCDAs.

Usually, when one calculating TFFs using the LCSR with traditional currents, twist-2, 3, 4 LCDAs will include. However, now only twist-2 LCDAs have been systematically investigated~\cite{Lepage:1979zb, Efremov:1979qk, Zhang:2020gaj, deMelo:2015yxk, Arthur:2010xf, Zhong:2014jla, Zhong:2014fma, Zhang:2017rwz, Zhong:2021epq}. In contrast, research on twist-3 and twist-4 LCDAs remains relatively insufficient~\cite{Han:2013zg, Lu:2006fr, Zhong:2016kuv}. Systematic numerical analyses have revealed key characteristics of twist contributions in $D$-meson decays: twist-4 effects account for less than 4\% LCDAs of the total sum rule results, while twist-3 contributions are numerically comparable to twist-2 contributions~\cite{Huang:2004su}. Specifically, in the energy region $q^2 < 10 ~ \text{GeV}^2$, twist-3 contributions remain significant: their magnitude is close to that of twist-2 terms, and provide a non-negligible supplement to TFFs of pion meson.

For the high twist LCDAs, we currently do not know its specific form. If all of them are taken into account at once, it will introduce large errors. Therefore, conducting targeted research on a certain LCDA is also something that people are doing. In this regard, a new form for calculating TFFs has been developed in LCSR. Specifically, by properly choosing the correlator, one can concentrate on different twist structures of pion LCDA within the LCSR~\cite{Zhong:2011jf}. There are two types of chiral correlators: right-handed and left-handed. The right-handed one cancels the twist-3 terms appropriately, so one only needs to consider the twist-2 and twist-4 LCDAs \cite{Huang:2008zg}, suppressing the influence of the twist-3 LCDAs, thus obtaining more precise LCSR predictions for TFFs \cite{Zuo:2006dk, Wu:2007vi, Wu:2009kq, Cheng:2017bzz, Zhou:2019jny, Ball:1998je, Ball:2006wn}. Conversely, the left-handed one exactly the terms of the twist-2 LCDA, thereby focusing solely on the dominant twist-3 LCDAs~\cite{Zhou:2004dq}. Based on the above analysis, it can be seen that the introduction of chiral currents not only simplifies the framework of the LCSR but also reduces the uncertainty of the results. In this paper, we choose to adopt these two chiral current correlators to calculate the TFF for comparison. Based on the previous analysis, the current research on twist-3 LCDA is not yet mature enough in terms of precision. To avoid introducing more additional errors, this paper only considers the influence of the leading order (LO), and the calculation of the twist-2 LCDA is up to next-to-leading orde(NLO) in order to enhance computational precision \cite{Khodjamirian:1999hb}.

Moreover, as among the most important nonperturbative parameters, the pion meson twist-2 and twist-3 LCDAs, which encapsulate the long-distance dynamics at lower energy scales, provide the dominant contribution to the calculation of the TFFs~\cite{Yang:2024ang}. Therefore, a detailed investigation of these LCDAs is crucial for improving the precision of TFF calculations. For the pion meson twist-2 LCDA, various predictions exist from different theoretical approaches, such as LQCD~\cite{RQCD:2019osh}, the DS model~\cite{Chang:2013pq}, and the AdS/QCD~\cite{Ahmady:2018muv}. Predictions for the pion meson twist-3 LCDAs have been provided by the light-front quark model (LFQM)~\cite{Arifi:2023uqc} and QCDSR~\cite{Huang:2004tp, Huang:2005av, Braun:1989iv}. In this work, to obtain endpoint behavior that better conforms to theoretical expectations, we will employ the light-cone harmonic oscillator model (LCHO) for the pion meson twist-2 and twist-3 LCDAs, determining its model-dependent parameters by constructing a set of three constraint equations.

This paper is organized as follows. In Sec.~\ref{Sec:2}, we present the calculation of the $D \to \pi$ TFF within the LCSR method using the right-handed and left-handed correlators, and also construct the pion twist-2 and twist-3 LCHO model. In Sec.~\ref{Sec:3}, we perform a detailed numerical analysis of the TFFs, the differential decay width, the branching fraction, and the CKM matrix element $|V_{cd}|$. Finally, Sec.~\ref{Sec:4} contains a summary.

\section{Theoretical framework}\label{Sec:2}
In order to study relevant physical observables, the full differential decay width distribution of $D \to \pi e\nu_e$ with respect to the squared momentum transfer $q^2$ and helicity angle $\theta_\ell$ can be written as $d\Gamma(D \to \pi e \nu_e)= a_{\theta_\ell}(q^2) + b_{\theta_\ell}(q^2) \cos\!\theta_\ell  + c_{\theta_\ell}(q^2) \cos^2\theta_\ell$. The $a_{\theta_\ell}(q^2)$, $b_{\theta_\ell}(q^2)$, $c_{\theta_\ell}(q^2)$ are the three $q^2$-dependent angular coefficient functions that can be expressed by two different TFFs $f^{D\pi}_{+}$~\cite{Becirevic:2016hea,Cui:2022zwm}. Meanwhile, the helicity angle $\theta_\ell$ is mainly coming form the angle between $\ell^-$ direction of flight and final-state meson momentum in dilepton rest frame. Thus, in the massless lepton limit, we can observe two interesting algebra relations for the angular functions $b_{\theta_\ell}(q^2) = 0$ and $a_{\theta_\ell}(q^2) + c_{\theta_\ell}(q^2) = 0$. By integrating over $q^2$ in entire physical region and utilizing the lifetime $\tau_D$, we can get the branching fraction $\mathcal{B}(D \to \pi e \nu_e) = \tau_D \int_0^{q_{\text{max}}^2} d\Gamma(D \to \pi e \nu_e)dq^2$
with $q_{\text{max}}^2 = (m_D - m_\pi)^2$.
Furthermore, we should carry out a calculation for the $D \to \pi$ TFFs. In the first place, to derive LCSR expressions for the $D \to \pi$ TFFs, we will take the right-chiral and left-chiral current correlator to calculate $f_{+}^{D\pi}(q^2)$.
For the next step, we can start from the vacuum to meson correlation function to derive the LCSR of the TFF. The detailed form can be written as
\begin{align}
\Pi_\mu(p, q) &= i \int d^4 x e^{iqx} \langle \pi(p) | T \{ j_\mu(x), j_{D}^\dagger(0) \} | 0 \rangle \nonumber \\
&= \Pi(q^2, (p + q)^2) p_\mu + \tilde{\Pi}(q^2, (p + q)^2) q_\mu.
\end{align}

Here, we employ two distinct chiral currents to compute the correlation functions: the first one is the right-handed chiral current, namely Scheme I: $ j_{\mu}^{\text{I}}(x)= \bar{{d}}(x)\gamma_{\mu}(1+\gamma_{5}) {c}(x) $ and $ j_{D}^{\dagger \text{I}}(0) = \bar{{c}}(0)i(1+\gamma_{5}){d}(0) $; The Scheme II: $ j_{\mu}^{\text{II}}(x)= j_{\mu}^{\text{I}}(x)$ and $ j_{D}^{\dagger \text{II}}(0)= \bar{{c}}(0)i(1-\gamma_{5}) {d}(0) $. By selecting the right-handed current correlator, we can eliminate the contributions from the twist-3 LCDAs, ensuring that the nonperturbative inputs in TFFs are primarily determined by the twist-2 LCDA. This right-handed current approach effectively compensates for our limited understanding of the twist-3 LCDA and the absence of corresponding $\mathcal{O}(\alpha_s)$ corrections~\cite{Wu:2025kdc}. Conversely, the use of the left-handed current correlator can isolate expressions that depend solely on the twist-3 LCDA, avoiding both the dominant contribution from the twist-2 LCDA and the minor effects of twist-4, and thus focus exclusively on twist-3 physics~\cite{Zhong:2011jf, Zhou:2004dq}. This facilitates the observation of twist-3 LCDAs properties and the elimination of other extraneous interference effects. Although the use of chiral currents may introduce additional uncertainties from scalar $(0^+)$ $D$-meson resonances, these can be absorbed into the corresponding hadronic dispersion integral through a proper choice of the continuum threshold~\cite{Zhou:2019jny}. In the paper, we adopt the above two choices of correlators to show whether the LCSRs under different correlators are consistent with each other.

After taking the $c$-quark propagators into the correlation function and making OPE, we can get the $D \to \pi$ TFF OPE expression. On the other hand, by inserting the complete intermediate states with the same quantum numbers as the current operator $j_D^\dagger(0)$ in the correlator and isolating the pole term of the lowest pseudoscalar $D$ meson, we can get the hadronic representation of the correlator. With the help of quark-hadron duality and introducing the effective threshold parameter $s_0$, the sum rules for $D \to \pi$ TFFs are obtained by making the Borel transformation in the variable $(p+q)^2 \to M^2$. The expressions for the TFFs obtained under the two different chiral current schemes are

\begin{align}
f_{+}^{D\pi(\text{I})}(q^{2}) &= \frac{e^{m_{D}^{2}/M^{2}}}{m_{D}^{2}f_{D}} \bigg[ F_{0}^{\text{I}}(q^{2}, M^{2}, s_{0})  \nonumber \\
& + \frac{\alpha_{s}C_{F}}{4\pi}F_{1}^{\text{I}}(q^{2}, M^{2}, s_{0}) \bigg],
\\
f_{+}^{D\pi(\text{II})}(q^{2}) &= \frac{e^{m_{D}^{2}/M^{2}}}{m_{D}^{2}f_{D}} F_{0}^{\text{II}}(q^{2}, M^{2}, s_{0}).
\end{align}
In which the $F_0^\text{I}$ and $F_0^\text{II}$ stand for the LO contributions, while $F_1^\text{I}$ represent the NLO contributions. Their respective expressions are given as follows:
\begin{widetext}
\begin{align}
F_{0}^{\text{I}}(q^{2}, M^{2}, s_{0}) &= m_{c}^{2}\,f_{\pi} \,\Bigg\{ \, \int_{\Delta}^{1} \,du \,\exp\left[ -\frac{m_{c}^{2} \,-\, q^{2}(1-u)}{uM^{2}} \right]\,
\left[ \frac{\phi_{2;\pi}(u, \mu)}{u} \,+ \,\frac{G_{4;\pi}(u)}{uM^{2}}\, - \, \frac{m_{c}^{2}\phi_{4;\pi}(u)}{4u^{3}M^{4}} \right]
\nonumber \\
& + \int_{0}^{1} d \nu \int D\alpha_{i} \frac{\theta(\alpha_{1} + \nu\alpha_{3} - \Delta)}{(\alpha_{1} + \nu\alpha_{3})^{2}M^{2}}
\, \exp\left[ {-\frac{m_{c}^{2}(1\,-\alpha_{1}\,-\nu\alpha_{3}) \, [q^{2}-(\alpha_{1}\,+\,\nu\alpha_{3}) m_{\pi}^{2}]}{(\alpha_{1}+\nu\alpha_{3})M^{2}}}\right]
\nonumber \\
& \times \left[ 2\Psi_{4;\pi}(\alpha_{i}) + 2\tilde{\Psi}_{4;\pi}(\alpha_{i}) - \Phi_{4;\pi}(\alpha_{i}) - \tilde{\Phi}_{4;\pi}(\alpha_{i}) \right] \Bigg\},
\\
F_{0}^{\text{II}}(q^{2}, M^{2}, s_{0})& = \mu_{\pi} m_{c} f_{\pi} \Bigg\{ \int_{\Delta}^{1} du \exp\left[ -\frac{m_{c}^{2} - q^{2}(1-u)}{uM^{2}} \right]
\left[ u\phi_{3;\pi}^{p}(u) + \frac{1}{6}\left( 2 + \frac{m_{c}^{2} + q^{2}}{uM^{2}} \right)\phi_{3;\pi}^{\sigma}(u) \right]
\nonumber \\
& - \frac{2f_{3\pi}}{f_{\pi}\mu_{\pi}}\, \int_{0}^{1}\, d\nu \, \int D\alpha_{i} \frac{\theta(\alpha_{1} + \nu\alpha_{3} - \Delta)}{(\alpha_{1} + \nu\alpha_{3})^{2}}
\exp\left[ -\frac{m_{c}^{2}\, - q^{2}(1 \,- \alpha_{1} \,- \nu\alpha_{3})}{(\alpha_{1} + \nu\alpha_{3})M^{2}} \right]\, \Phi_{3;\pi}(\alpha_{i})
\nonumber \\
& \times \left[ 1 - \frac{m_{c}^{2} - q^{2}}{(\alpha_{1} + \nu\alpha_{3})M^{2}} \right] \Bigg\},
\\
 F_{1}^{\text{I}}\left(q^{2}, M^{2}, s\right)& = \frac{f_{\pi}}{\pi} \int_{m_{c}^{2}}^{s_{0}} ds e^{-s/M^{2}}
\int_{0}^{1} du \left[ \text{Ims} \, T_{1}\left(q^{2}, s, u\right) \phi_{2;\pi}(u) \right],
\end{align}
\end{widetext}
where the parameters are defined as $\mu_\pi = m_\pi^2/(m_u + m_d)$, $\Delta = (m_c^2 - q^2)/(s_0 - q^2)$, $G_{4;\pi}(u) = -\int_0^u dv \,\psi_{4;\pi}(v)$, and $\mathcal{D}\alpha_i = d\alpha_1 d\alpha_2 d\alpha_3 \,\delta(1 - \alpha_1 - \alpha_2 - \alpha_3)$. The specific form of $\operatorname{Ims} T_{1}(q^2, s, u)$ is provided in Ref.~\cite{Duplancic:2008ix}. Here, $\phi_{2;\pi}(u, \mu)$ denotes the twist-2 LCDA of the pion, while $\phi^p_{3;\pi}(u,\mu)$ and $\phi^\sigma_{3;\pi}(u,\mu)$ correspond to its twist-3 LCDAs. For twist-4 contributions, we introduce the pion LCDAs $\phi_{4;\pi}(u)$, $\psi_{4;\pi}(u)$, $\Psi_{4;\pi}(\alpha_i)$, $\tilde{\Psi}_{4;\pi}(\alpha_i)$, $\Phi_{4;\pi}(\alpha_i)$, and $\tilde{\Phi}_{4;\pi}(\alpha_i)$.

Furthermore, the twist-2 and twist-3 LCDAs of the pion meson are the most significant non-perturbative parameters contributing to the calculation of the TFFs. In this study, the LCHO model is employed to analyze them. This model is constructed based on the BHL framework~\cite{Wu:2010zc, Wu:2011gf} and can link the equal-time wave function in the rest frame with the light-cone wave function (LCWF) in the moving frame and incorporates the Wigner-Melosh rotation effect. Specifically, the twist-2 and twist-3 LCDAs of the pion meson, along with their corresponding LCWFs, are defined as follows~\cite{Zhong:2021epq}
\begin{align}
\phi_{i;\pi}(x,\mu) = \mathcal{N}_i \int_{|\mathbf{k}_\perp|^2 \leq \mu^2} \frac{d^2 \mathbf{k}_\perp}{16\pi^3} \Psi_{i;\pi}(x, \mathbf{k}_\perp),
\label{WF and DA relation}
\end{align}
where, $\mathbf{k}_\perp$ denotes the pion meson transverse momentum, and the index $i=(2,3)$ corresponds to the twist-2 and twist-3 LCDAs, respectively. The normalization coefficients are $\mathcal{N}_2 = 2\sqrt{6}/f_{\pi}$ and $\mathcal{N}_3 = 1$, while $\Psi_{i;\pi}(x,\mathbf{k}_\perp)$ represent the wave functions for the twist-2 and twist-3 pion states, respectively. Under the BHL framework, the corresponding pion meson twist-2 LCWFs can be derived from the relationship between the spin wave function and the spatial wave function.
\begin{equation}
\Psi_{2;\pi}(x, \mathbf{k}_\perp) = \sum_{\lambda_1 \lambda_2} \chi_{2;\pi}^{\lambda_1 \lambda_2}(x, \mathbf{k}_\perp) \Psi_{2;\pi}^R(x, \mathbf{k}_\perp),
\end{equation}

Where the $\chi_{2;\pi}^{\lambda_1\lambda_2}(x, \mathbf{k}_\perp) $ denotes the spin-space WF based on the Wigner-Melosh rotation, with its different $\lambda_1\lambda_2$ forms given in Refs.~\cite{Cao:1997hw, Wu:2005kq}. Meanwhile, the spatial wave function $\Psi_{2;\pi}^{\text{R}}(x, \mathbf{k}_\perp)$ is defined as follows:
\begin{equation}
\Psi_{2;\pi}^{\text{R}}(x, \mathbf{k}_\perp) = A_{2;\pi} \varphi_{2;\pi}(x) \exp\left[ -\frac{\mathbf{k}_\perp^2 + m_q^2}{8 \beta_{2;\pi}^2 x \bar{x}} \right],
\end{equation}
where $A_{2;\pi}$ the normalization constant. The $\mathbf{k}_\perp$-dependence part of the spatial WF $\Psi_{2;\pi}^R(x, \mathbf{k}_\perp)$ arises from the approximate bound-state solution in the quark model for pion and determines the WF's transverse distribution via the harmonious parameter $\beta_{2;\pi}$, while $\varphi_{2;\pi}(x)$ dominates the WF's longitudinal distribution.

After integrating over the transverse momentum $\bf k_{\bot}$, we obtain the leading-twist LCDA $\phi_{2;\pi}(x, \mu)$ of the pion meson, which is expressed as
\begin{equation}
\begin{split}
& \phi_{2;\pi}(x,\mu)  = \frac{\sqrt{3} A_{2;\pi} m_q \beta_{2;\pi}}{2\pi^{3/2} f_\pi} \sqrt{x \bar{x}} \varphi_{2;\pi}(x) \\
& \qquad \times \left\{ \text{Erf}\left[ \sqrt{\frac{m_q^2 + \mu^2}{8 \beta_{2;\pi}^2 x \bar{x}}} \right] - \text{Erf}\left[ \sqrt{\frac{m_q^2}{8 \beta_{2;\pi}^2 x \bar{x}}} \right] \right\},
\end{split}
\end{equation}
here, $\bar{x} = 1 - x$, $\varphi_{2;\pi}(x) = 1 + B \times C_2^{3/2}(2x - 1)$, $C_{2}^{3/2}$ denotes a Gegenbauer polynomial, and $\text{Erf}(x) = 2 \int_0^x e^{-t^2} dx / \sqrt{\pi}$ stands for the error function. We fix the quark mass at $m_q = 250$ MeV in the present work. Evidently, the behavior of $\phi_{2;\pi}(x,\mu)$ is governed by the unknown model parameters $A_{2;\pi}$, $\beta_{2;\pi}$, and $B$, and their determination requires additional constraints:
\begin{itemize}
    \item Normalization condition of the twist-2 LCDA for the pion:
    \begin{equation}
        \int_0^1 dx \int \frac{d^2 \mathbf{k}_\perp}{16\pi^3} \Psi_{2;\pi}(x, \mathbf{k}_\perp) = \frac{f_\pi}{2\sqrt{6}}.
    \end{equation}
    \item Average value of the squared transverse momentum $\langle \mathbf{k}_\perp^2 \rangle_{2;\pi}$:
    \begin{equation}
        \langle \mathbf{k}_\perp^2 \rangle_{2;\pi} = \frac{\int dx \, d^2 \mathbf{k}_\perp \, \mathbf{k}_\perp^2 | \Psi_{2;\pi}(x, \mathbf{k}_\perp) |^2}{\int dx \, d^2 \mathbf{k}_\perp | \Psi_{2;\pi}(x, \mathbf{k}_\perp) |^2},
    \end{equation}
    where $\langle \mathbf{k}_\perp^2 \rangle_{2;\pi}^{1/2} = 0.350\ \text{GeV}$~\cite{Guo:1991eb, Zhong:2011rg}.
  \item The Gegenbauer moments $a_n^{2;\pi}(\mu)$ can be derived by the following way:
    \begin{equation}
        a_n^{2;\pi}(\mu) = \frac{\int_0^1 dx \, \phi_{2;\pi}(x,\mu) \, C_n^{3/2}(\xi)}{\int_0^1 dx \, 6x\bar{x} [ C_n^{3/2}(\xi) ]^2}.
    \end{equation}
    with $\xi=2x-1$. For $n=2$, we adopt the $a_2^{2;\pi}(\mu) = 0.101 \pm 0.023$ of LQCD at $\mu = 2\ \text{GeV}$~\cite{RQCD:2019osh}.
\end{itemize}

Meanwhile, the two twist-3 LCDAs of pion meson can also be related to their WF by using the formula
\begin{align}
\Psi_{3;\pi}^p (x, \mathbf{k}_\perp) &= [1 + B_{3;\pi}^p C_2^{1/2}(2x - 1)] \frac{A_{3;\pi}^p}{x(1-x)} \nonumber \\
& \times \exp\left[ -\frac{m_q^2 + \mathbf{k}_\perp^2}{8\beta_{3;\pi}^{p2} x(1-x)} \right],
\\
\Psi_{3;\pi}^{\sigma}(x, \mathbf{k}_\perp) &= [1 + B_{3;\pi}^{\sigma} C_2^{3/2}(2x - 1)] \frac{A_{3;\pi}^\sigma}{x(1-x)}
\nonumber \\
& \times \exp\left[ -\frac{m_q^2 + \mathbf{k}_\perp^2}{8\beta_{3;\pi}^{\sigma 2} x(1-x)}\right].
\end{align}
Using the relation between the LCDA and the WF, we obtain~\cite{Chen:2024xdd}
\begin{align}
&\phi_{3;\pi}^{p}(x, \mu)  = \frac{A_{3;\pi}^{p} \beta_{3;\pi}^{p2}}{2\pi^2}  [1 + B_{3;\pi}^p C_2^{1/2}(2x-1)]
\nonumber \\
& \quad \times \exp\left[ -\frac{m_q^2}{8 \beta_{3;\pi}^{p2} x \bar{x}} \right] \left[ 1 - \exp\left( -\frac{\mu^2}{8 \beta_{3;\pi}^{p2} x \bar{x}} \right) \right],
\\
&\phi_{3;\pi}^{\sigma}(x, \mu)  = \frac{A_{3;\pi}^{\sigma} \beta_{3;\pi}^{\sigma 2}}{2\pi^2} [ 1 + B_{3;\pi}^\sigma C_2^{3/2}(2x-1) ]
\nonumber \\
& \quad \times \exp\left[ -\frac{m_q^2}{8 \beta_{3;\pi}^{\sigma 2} x \bar{x}} \right] \left[ 1 - \exp\left( -\frac{\mu^2}{8 \beta_{3;\pi}^{\sigma 2} x \bar{x}} \right) \right].
\end{align}
The parameters $A_{3;\pi}^{p,\sigma}$, $B_{3;\pi}^{p,\sigma}$, and $\beta_{3;\pi}^{p,\sigma}$ can similarly be determined by the following conditions.
\begin{itemize}
    \item Normalization condition of the WF
    \begin{equation}
        \int_{0}^{1} dx \int_{|\mathbf{k}_\perp|<\mu} \frac{d^2\mathbf{k}_\perp}{16\pi^3} \Psi_{p,\sigma}(x,\mathbf{k}_\perp) = 1
    \end{equation}
    \item Average value of the squared transverse momentum $\langle \mathbf{k}_\perp^2 \rangle_{p,\sigma}$:
    \begin{equation}
        \langle \mathbf{k}_\perp^2 \rangle_{p,\sigma} = \frac{\displaystyle\int dx \, d^2\mathbf{k}_\perp \, \mathbf{k}_\perp^2 \left| \Psi_{p,\sigma}(x, \mathbf{k}_\perp) \right|^2}{\displaystyle\int dx \, d^2\mathbf{k}_\perp \left| \Psi_{p,\sigma}(x, \mathbf{k}_\perp) \right|^2}
    \end{equation}
    where $\langle \mathbf{k}_\perp^2 \rangle_{p,\sigma}^{1/2} = 0.350\ \text{GeV}$~\cite{Zhong:2011jf}.
    \item The first two Gegenbauer moments $a_{p}^{n}(\mu)$ and $a_{\sigma}^{n}(\mu)$ can be derived by the following way.
    \begin{align}
        &a_{p}^{n}(\mu) = \frac{\displaystyle\int_{0}^{1} dx \, \phi_{3;\pi}^{p}(x,\mu) \, C_{n}^{1/2}(\xi)}{\displaystyle\int_{0}^{1} dx \, [ C_{n}^{1/2}(\xi) ]^2}, \\
        &a_{\sigma}^{n}(\mu) = \frac{\displaystyle\int_{0}^{1} dx \, \phi_{3;\pi}^{\sigma}(x,\mu) \, C_{n}^{3/2}(\xi)}{\displaystyle\int_{0}^{1} dx \, 6x(1 - x) [ C_{n}^{3/2}(\xi) ]^2},
    \end{align}
      When $n=2$, we adopt the determined values $a_p^{2}(\mu_0) = 0.44296$ and $a_{\sigma}^{2}(\mu_0) = 0.01658$ at the initial scale $\mu_0 = 1\ \text{GeV}$~\cite{Nam:2006mb}.
\end{itemize}

Through the above discussion, not only can we determine the model parameters of the twist-2 and twist-3 LCDAs, but also we have laid a theoretical foundation for calculating the relevant observables of the semileptonic decay process $D\to \pi$.

\section{NUMERICAL ANALYSIS}\label{Sec:3}
In order to perform the phenomenological analysis, we adopt following basic input parameters from PDG'24~\cite{ParticleDataGroup:2024cfk}. The mass of $c$-quark $ m_c(\bar{m}_c) = 1.27 \pm 0.02 \ \text{GeV}$. The masses of the $D$ and pion mesons are $ m_{D^0} = 1864.84 \pm 0.05 \, \text{MeV} $, $ m_{D^+} = 1869.66 \pm 0.05 \, \text{MeV} $, and $ m_\pi = 139.57~ \text{MeV} $. The decay constant is $f_D = 212^{+0.7}_{-0.7} \, \text{MeV}$ and $f_\pi = 130.2^{+0.8}_{-0.8} \, \text{MeV}$. For the $D \to \pi$ decay process, we first define the typical process energy scale as $\mu_k = \sqrt{m_D^2 - m_c^2} \approx 1.4 \, \text{GeV}$.

Based on the above constraints, we can further determine the model parameters of the twist-2 and twist-3 components in the pion meson LCDAs. However, in order to obtain the LCDA parameters at a specific energy scale $\mu_k$, we need to evolve the parameters at the initial energy scale $\mu_0$ to the target energy scale through the renormalization group equation (REG). It is particularly important to note that the REG governing the Gegenbauer moments of the twist-2 and twist-3 LCDAs are not the same. This will be discussed below.

For the twist-2 case, the REG for its energy spectrum evolution is defined as follows.
\begin{equation}
E_n(\mu_{k}, \mu_0) = \left[ \frac{\alpha_s(\mu_{k})}{\alpha_s(\mu_0)} \right]^{\gamma_n^{(0)}/(2\beta_0)},
\end{equation}
here, $a_n^{2; \pi}(\mu_{k}) = a_n^{2; \pi}(\mu_0) E_n(\mu_{k}, \mu_0)$, The $\gamma_n^{(0)}$ denotes the LO anomalous dimension. The definition is as follows
\begin{equation}
\gamma_n^{(0)} = 8C_F \!\left[ \psi(n+2) +\! \gamma_E - \!\frac{3}{4} -\! \frac{1}{2(n+1)(n+2)} \right],
\end{equation}
with $C_F = 4/3$.
Similarly, the energy spectrum evolution in the twist-3 case is described by the following REG.
\begin{equation}
a_{p(\sigma)}^{n}(\mu_{k}) = a_{p(\sigma)}^{n}(\mu_0) \left( \frac{\alpha_s(\mu_{k})}{\alpha_s(\mu_0)} \right)^{\gamma_n / \beta_0},
\end{equation}
where $\gamma_n = C_F \left( 1 - \frac{2}{(n+1)(n+2)} + 4 \sum_{m=2}^{n+1} \frac{1}{m} \right)$ and $\beta_0 = (11N_c - 2N_f)/3$.

\begin{figure}[h]
\begin{center}
\includegraphics[width=0.435\textwidth]{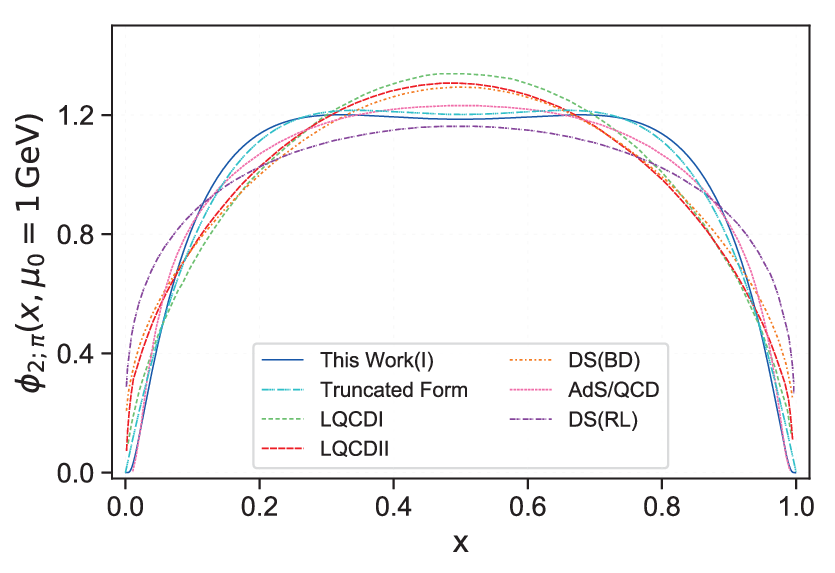}
\caption{The pion twist-2 LCDA at initial scale $\mu_0 = 1$ GeV. For comparison, we also include results from truncated form, LQCD \cite{RQCD:2019osh}, DS model (BD and RL schemes) \cite{Chang:2013pq}, and AdS/QCD model \cite{Ahmady:2018muv}.}
\label{Fig:DA-R}
\end{center}
\end{figure}

\begin{figure}[h]
\begin{center}
\includegraphics[width=0.435\textwidth]{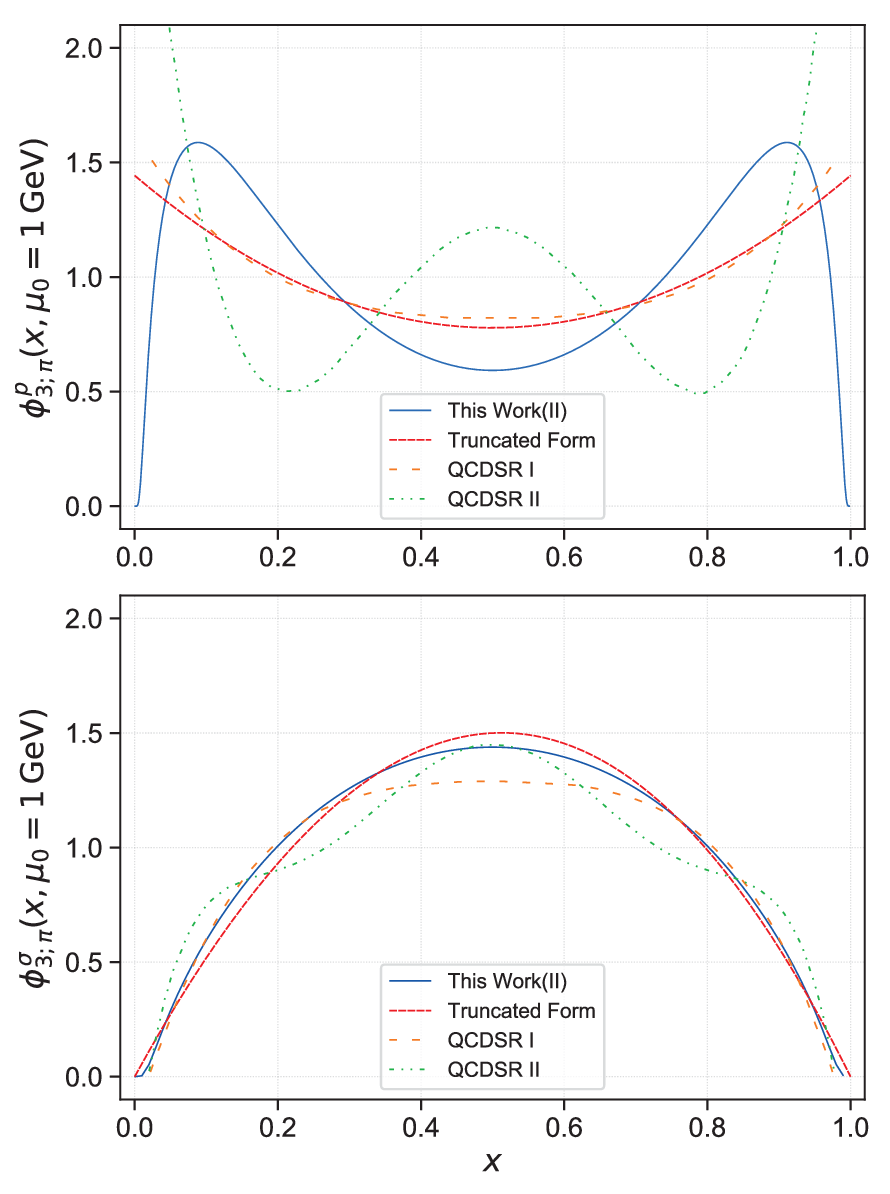}
\caption{The pion twist-3 LCDAs $\phi_{3;\pi}^p(x,\mu_0)$ and $\phi_{3;\pi}^\sigma(x,\mu_0)$ at initial scale $\mu_0$ that adopted in the Scheme-II. As a comparison, the truncated from, QCDSR-I~\cite{Ball:1998je} and QCDSR-II~\cite{Braun:1989iv} are also present here.}
\label{Fig:DA-L}
\end{center}
\end{figure}

After the above calculations, we obtained the numerical results of the twist-2 and twist-3 LCDAs model parameters for the pion at the energy scales of $\mu_0 = 1\ \mathrm{GeV}$ and $\mu = 1.4\ \mathrm{GeV}$, and summarized them in Table~\ref{tab:11}. Using the data in Table~\ref{tab:11}, \add{we further present the behavior curves of twist-2 and twist-3 LCDAs at initial scale $\mu_0$ in Fig.~\ref{Fig:DA-R} and~\ref{Fig:DA-L}, respectively. As a comparison, in Fig.~\ref{Fig:DA-R}, the predictions from truncated form, LQCD model \cite{RQCD:2019osh}, DS model \cite{Chang:2013pq}, AdS/QCD model \cite{Ahmady:2018muv} theoretical methods are also given, and in Fig.~\ref{Fig:DA-L}, the prediction curves of truncated form, QCDSR-I~\cite{Ball:1998je} and QCDSR-II~\cite{Braun:1989iv} are given, respectively. Here the truncated form mainly comes from the Gegenbauer polynomial expansion truncated up to $n=2$ order which is more tending to the real behavior of pion LCDA, that is also taken for the lattice QCD group. In this work, we take the central value of the second order Gegenbauer moment $a_{2;\pi}(\mu_0)$ from the lattice results~\cite{RQCD:2019osh} and $a_{3;\pi}^{p,\sigma}(\mu_0)$ from flavor SU(3) symmetry breaking~\cite{Nam:2006mb}.
The direct comparison shows that our prediction results are consistent with the above theoretical results, exhibiting a single-peak feature and symmetry.}
\begin{table}[htbp]
  \centering
  \caption{Parameters of the pion twist-2 and 3 LCDAs within LCHO model at different scales}
  \label{tab:11}
  \renewcommand{\arraystretch}{1.3}
  \small
  \begin{tabular}{lllll}
  \hline
  Scale~~~~~~~~    &    LCDA ~~~~~~~                      & $A$~~~~~~~~    &  $B$~~~~~~~~   &  $\beta$ \\ \hline
           & $\phi_{2;\pi}(x)$        & 26.33  &  0.08 &  0.58 \\
  $\mu_0$  & $\phi_{3;\pi}^p(x)$      & 101.5  &  0.93  &  0.54  \\
           & $\phi_{3;\pi}^\sigma(x)$ & 202.6  &  $-0.10$ &  0.39  \\ \hline
           & $\phi_{2;\pi}(x)$        & 24.69  &  0.08 &  0.58  \\
  $\mu_k$  & $\phi_{3;\pi}^p(x)$      & 104.6  &  0.87  &  0.53  \\
           & $\phi_{3;\pi}^\sigma(x)$ & 203.3  &  $-0.11$ &  0.39  \\
    \hline
  \end{tabular}
\end{table}

Furthermore, when calculating the TFFs for the $D \to \pi$ decay process, two important input parameters need to be determined, namely the continuous threshold $s_0$ and the Borel parameter $M^2$. These parameters can be determined based on the self-consistency criterion of the QCDSR~\cite{Tian:2023vbh}. The specific criteria are as follows:
\begin{itemize}
    \item The continuum contributions are less than 30\% to the total results;
    \item The contributions from higher-twist LCDAs are less than 5\%;
    \item Within the Borel window, the changes of TFFs does not exceed 10\%;
    \item The continuum threshold $s_0$ should be closer to the squared mass of the first excited state $D$-meson.
\end{itemize}

Based on the above criteria, the continuum threshold $s_0$ and Borel parameter $M^2$ corresponding to the two methods are determined as follows: for Scheme I, $s_0 = 6.3(1)~\text{GeV}^2 $ and $M^2 = 15(2)~ \text{GeV}^2$; for Scheme II, $ s_0 = 6.0(1) ~\text{GeV}^2$ and $M^2 = 4.0(5)~ \text{GeV}^2$.

\begin{figure}[h]
\begin{center}
\includegraphics[width=0.435\textwidth]{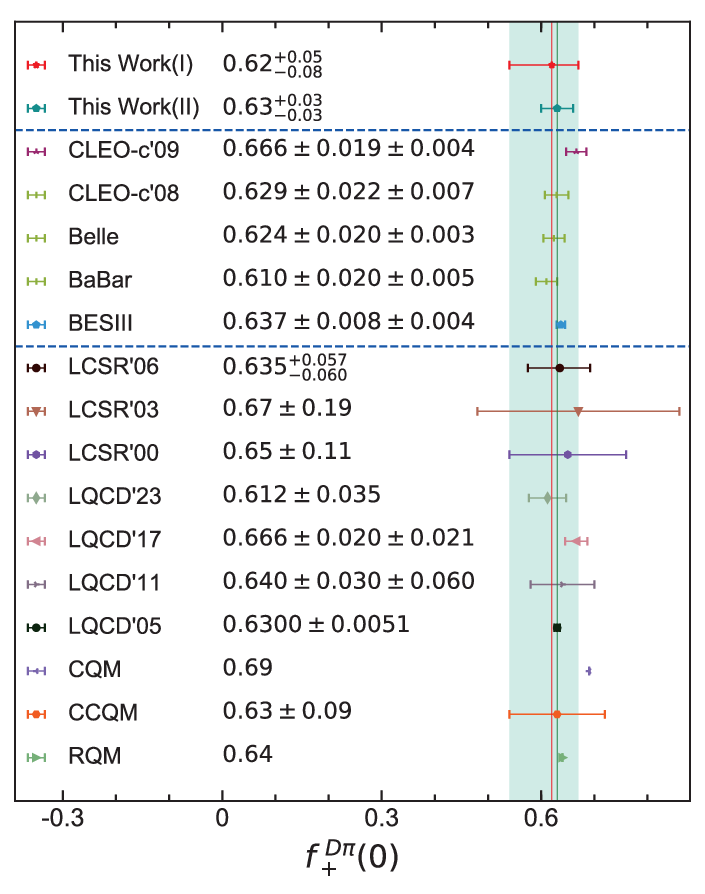}

\caption{The $D\to \pi$ TFF $ f_+^{D\pi}(0) $ at large recoil point. As a comparison, we also present other experimental and theoretical predictions.}
\label{Fig:TFF0}
\end{center}
\end{figure}

Subsequently, we calculated the values of the $D \to \pi$ TFFs in the large recoil region $f_+^{D\pi}(0)$ for both methods. The results were numerically consistent: the central values of the two methods differed by only approximately 1.3\%. This result is shown in Fig.~\ref{Fig:TFF0}. To facilitate comparison, Fig.~\ref{Fig:TFF0} also presents the experimental measurements and theoretical predictions from the experimental groups CLEO-c'09~\cite{ CLEO:2009svp}, CLEO-c'08~\cite{CLEO:2007ntr}, Belle~\cite{Belle:2006idb}, BaBar~\cite{BaBar:2014xzf}, and BESIII~\cite{BESIII:2015tql}, as well as the theoretical groups LCSR'06~\cite{Wu:2006rd}, LCSR'03~\cite{Wang:2002zba}, LCSR'00~\cite{Khodjamirian:2000ds}, LQCD'23~\cite{FermilabLattice:2022gku}, LQCD'17~\cite{Lubicz:2017syv}, LQCD'11~\cite{Na:2011mc}, LQCD'05\cite{FermilabLattice:2004ncd}, CQM~\cite{Melikhov:2000yu}, CCQM~\cite{Soni:2018adu}, and RQM~\cite{Faustov:2019mqr}. The analysis shows that the obtained values are highly consistent with the experimental results from CLEO'08, Belle, and BESIII, as well as the theoretical predictions from LCSR'06, LQCD'05, and CCQM. Within the error range, our calculation results are in agreement with the above experimental and theoretical predictions and are consistent with the vast majority of the predictions within their respective uncertainties. The consistency between the current experimental and theoretical results collectively supports the reliability of the current theoretical predictions and experimental data.

\begin{figure}[h]
\begin{center}
\includegraphics[width=0.435\textwidth]{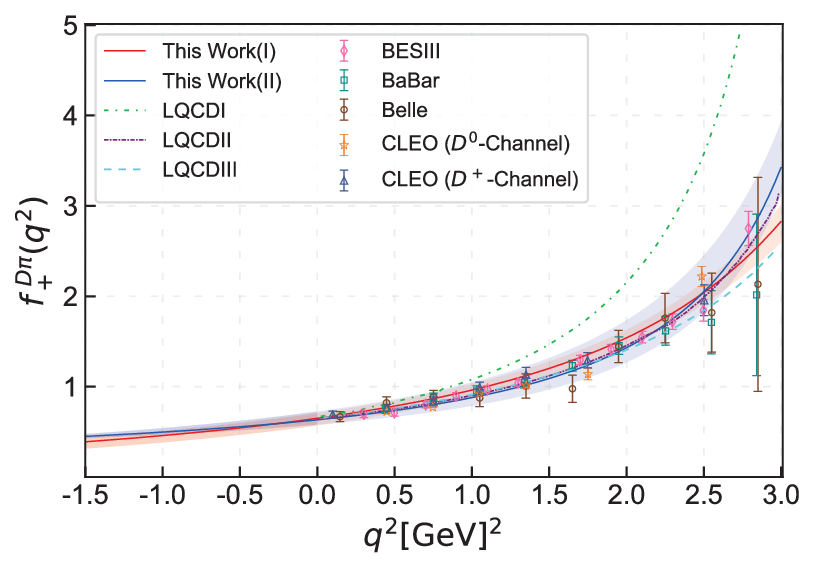}
\caption{The TFF $f_{+}^{D\pi}(q^2)$ versus $q^2$ in the whole physical region for Schemes I and II. As a comparison, we also give other experimental and theoretical predictions for comparison.}
\label{Fig:TFF}
\end{center}
\end{figure}

The physically allowable range for the $D\to \pi $ TFF is $0 \leq q^2 \leq q^2_{\text{max}} = (m_D - m_\pi)^2 \sim 3 \text{ GeV}^2$. Theoretically, since the $D$-meson mass is not sufficiently large, a hard scale is absent, and thus the LCSR approach for $D \to \pi $ TFFs is reliable for calculations in the region $q^2 \leq 0\text{ GeV}^2$. One can then extrapolate it to whole $q^2$-regions via a rapidly $z(q^2, t)$ converging the simplified series expansion (SSE), i.e. the TFF is expand as \cite{Bharucha:2015bzk}
\begin{equation}
f_{+}^{D\pi}(q^2) = \frac{1}{1 - q^2 / m_D^2} \sum_{k=0,1,2} \beta_k z^k(q^2, t_0)
\end{equation}
where $\beta_k$ are real coefficients and $z(q^2, t)$ is the function,
\begin{equation}
z^k(q^2, t_0) = \frac{\sqrt{t_+ - q^2} - \sqrt{t_+ - t_0}}{\sqrt{t_+ - q^2} + \sqrt{t_+ - t_0}},
\end{equation}
with $t_\pm = (m_D \pm m_\pi)^2$ and $t_0 =t_\pm \left(1 - \sqrt{1 - t_- / t_+}\right)$. The SSE method possesses superior merit, which keeps the analytic structure correct in the complex plane and ensures the appropriate scaling, $f_{+}^{D\pi}(q^2) \sim 1 / q^2$ at large $q^2$. And the quality of fit $\Delta$ is devoted to take stock of the resultant of extrapolation, which is defined as
\begin{equation}
\Delta = \frac{\sum_i \left| F_i(t) - F_i^{\text{fit}}(t) \right|}{\sum_i \left| F_i(t) \right|} \times 100.
\end{equation}
After making an extrapolation for the TFF $ f_{+}^{D\pi}(q^2) $ to the whole physical $ q^2 $-regions, the behaviors of $D \to \pi$ TFFs in the whole physical region with respect to squared momentum transfer are given in Fig.~\ref{Fig:TFF}. In the Fig.~\ref{Fig:TFF}, the darker bands are the LCSR results of our prediction, while the lighter bands are the SSE predictions. For comparison, we also presented experimental results from the collaboration of Belle~\cite{Belle:2006idb}, BaBar \cite{BaBar:2014xzf}, CLEO \cite{CLEO:2009svp} and BESIII \cite{BESIII:2015tql}, as well as theoretical predictions from LQCDI~\cite{FermilabLattice:2004ncd}, LQCDII~\cite{FermilabLattice:2022gku} and LQCDIII~\cite{DiVita:2010mlb}. Within the error bars, our two curves show reasonable agreement with the existing experimental and theoretical results. This consistency further validates the reliability and feasibility of the calculation framework employing both right-handed and left-handed correlation functions. In addition, the results obtained by incorporating NLO correction into the Scheme I method are similar to those obtained by direct Scheme II calculation, which not only confirms the effectiveness of the $O(\alpha_s)$ correction scheme implemented in Scheme I, but also improves the overall computational accuracy.

\begin{figure}[h]
\begin{center}
\includegraphics[width=0.435\textwidth]{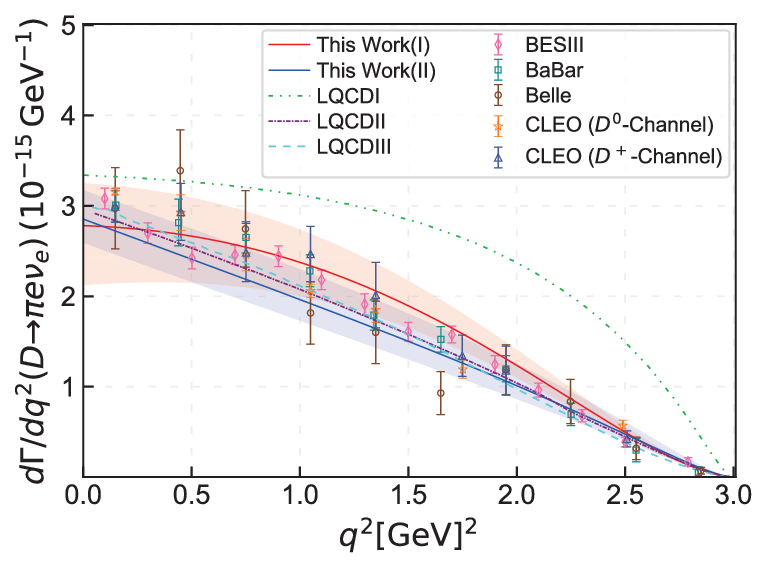}
\caption{The differential decay widths for $D \to \pi e\nu_e$ for schemes I and II. The result of the Belle~\cite{Belle:2006idb}, BESIII~\cite{BESIII:2015tql}, BaBar~\cite{BaBar:2014xzf}, and CLEO~\cite{CLEO:2009svp} Collaborations and the LQCD~\cite{FermilabLattice:2004ncd, FermilabLattice:2022gku, DiVita:2010mlb} prediction are presented as a comparison.}
\label{Fig:decaywidth}
\end{center}
\end{figure}

Subsequently, we can calculate the decay width of the semileptonic decay $D \to \pi$. The results are shown in Fig.~\ref{Fig:decaywidth}. For comparison, the measurement results from experiments such as Belle~\cite{Belle:2006idb}, BESIII~\cite{BESIII:2015tql}, BaBar~\cite{BaBar:2014xzf}, and CLEO ~\cite{CLEO:2009svp}, as well as the theoretical prediction of LQCDI~\cite{FermilabLattice:2004ncd}, LQCDII~\cite{FermilabLattice:2022gku} and LQCDIII~\cite{DiVita:2010mlb}, are also presented in the figure. From the figure, it can be seen that although the curves given by the two methods are slightly different, both fall completely within the experimental data range. The results of Scheme I and Scheme II obtained in this paper are all within the experimental error band throughout the $q^2$ range and are close to the theoretical  results from LQCDII and LQCDIII. Overall, our calculation results are reasonable agreement with the existing experimental and theoretical data, and the two schemes are mutually consistent, thereby verifying the reliability of this study.

Furthermore, according to the data provided by PDG'24, the decay lifetime of the $D$ meson is $\tau_{D^0} = 0.4103 \ \text{ps}$ and $\tau_{D^+} = 1.033 \ \text{ps}$, and the CKM matrix element $|V_{cd}| = 0.221(4)$~\cite{ParticleDataGroup:2024cfk}. Based on these parameters, we calculated the decay branching fractions of $D^{(0,+)} \to \pi^{(-,0)} \ell \nu_{\ell}$, and the results are listed in Table~\ref{tab:fpi_values}. The error of our prediction comes from all input parameters. To make a comparison, we also present the PDG~\cite{ParticleDataGroup:2024cfk}, Belle~\cite{Belle:2006idb}, BESIII~\cite{BESIII:2015tql}, BES~\cite{BES:2004rav}, CLEO~\cite{CLEO:2009svp}, LQCDI~\cite{FermilabLattice:2004ncd}, LQCDII~\cite{FermilabLattice:2022gku} and LQCDIII~\cite{DiVita:2010mlb} predictions. The detailed analysis and discussion are as follows. For the $D^0 \to \pi^- e^+ \nu_e$ decay channel, the result of Scheme I is in excellent agreement with the theoretical predictions from LQCD, while the result of Scheme II shows good consistency with the experimental measurements from Belle and CLEO. Similarly, for the $D^+ \to \pi^0 e^+ \nu_e$ decay channel, the result of Scheme I nearly coincides with the CLEO measurement, and the result of Scheme II is also in high agreement with the BESIII measurement.

Although there are differences between the results of Scheme I and Scheme II, the central values of both schemes fall within the PDG error range, and the two sets of results have significant overlap within their respective error intervals. This overlap indicates that the two methods have good consistency under the corresponding uncertainties. Overall, the results obtained by the two schemes are consistent with the experimental results of Belle, CLEO, BESIII, and the theoretical LQCD. This not only verifies the reliability of the LCSR framework adopted in this paper but also provides a useful theoretical reference for research in the field of charmed physics to a certain extent.

\begin{table}[htbp]
\renewcommand{\arraystretch}{1.4}
\footnotesize
\centering
\caption{The predictions of the branching fractions within uncertainties (in unit: $10^{-2}$) for Scheme I and II cases. Meanwhile, the theoretical and experimental results from other groups are also given as a comparison.}
\label{tab:fpi_values}
\begin{tabular}{l@{\hspace{0.6em}} l @{\hspace{1.2em}} l}
    \hline
    Reference & $\mathcal{B}(D^0 \to \pi^- e^+ \nu_e)$  & $\mathcal{B}(D^+ \to \pi^0 e^+ \nu_e)$ \\
    \hline
    This Work (I) & $0.31^{+0.05}_{-0.05}$ & $0.39^{+0.06}_{-0.06}$ \\
    This Work (II) & $0.27^{+0.05}_{-0.03}$ & $0.34^{+0.06}_{-0.04}$ \\
    PDG~\cite{ParticleDataGroup:2024cfk} & $0.291 \pm 0.004$ & $0.372 \pm 0.017$ \\
    BESIII \cite{BESIII:2015tql} & $0.295 \pm 0.004 \pm 0.003$ & $0.363 \pm 0.008 \pm 0.005$ \\
    BES \cite{BES:2004rav} & $0.330 \pm 0.130 \pm 0.030$ & $\dots$ \\
    CLEO \cite{CLEO:2009svp} & $0.288 \pm 0.008 \pm 0.003$ & $0.405 \pm 0.016 \pm 0.009$ \\
    Belle \cite{Belle:2006idb} & $0.279 \pm 0.027 \pm 0.016$ & $\dots$  \\
    LQCD \cite{FermilabLattice:2004ncd} & $0.316 \pm 0.025 \pm 0.062$ & $\dots$ \\
\hline
\end{tabular}
\end{table}

\begin{figure}[h]
\begin{center}
\includegraphics[width=0.435\textwidth]{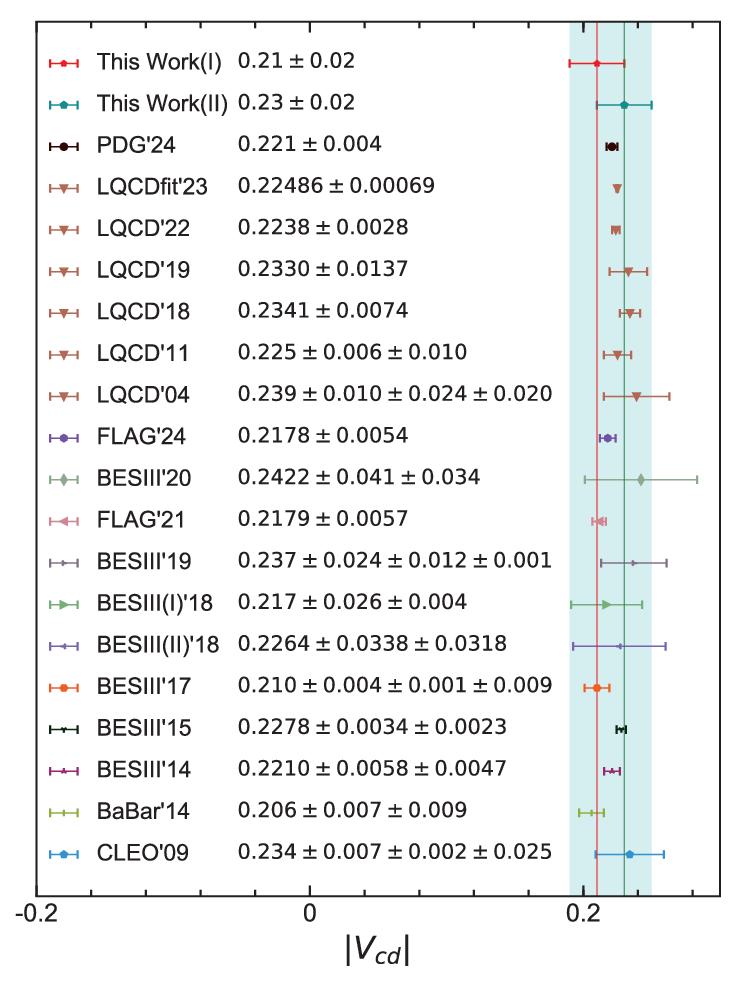}

\caption{The prediction of $|V_{cd}|$ from \ $D^0 \to \pi^- e^+ \nu_e$ within uncertainties for Scheme I and II cases, including other results of the experimental groups and the theoretical groups.}
\label{Fig:Vcd}
\end{center}
\end{figure}

In addition, due to the significant differences in $|V_{cd}|$ values derived from different decay modes, we attempt to provide a theoretical prediction for it through the $D^0 \to \pi^- e^+ \nu_e$ process using the following formula:
\begin{equation}
|V_{cd}| = \sqrt{\frac{\mathcal{B}_{\text{Exp}}(D^0 \to \pi^- e^+ \nu_e)}{\tau_{D^0} A}}
\end{equation}
here $ A $ represents the differential decay width independent of $|V_{cd}|$.
\begin{equation}
A = \frac{1}{|V_{cd}|^2} \int_{0}^{q_{\text{max}}^2} \frac{d\Gamma\left(D^0 \to \pi^- e^+ \nu_e\right)}{dq^2}
\end{equation}
Then we can obtain the CKM matrix element $|V_{cd}|$ from the branching fractions $\mathcal{B}(D^0 \to \pi^- e^+ \nu_e)$ in the PDG'24~\cite{ParticleDataGroup:2024cfk}. Our predictions are in Fig.~\ref{Fig:Vcd}, where the errors are caused by all the mentioned error sources and the PDG errors for the branching fractions and the decay lifetimes. For ease of comparison, we collected the CKM matrix element $|V_{cd}|$ results from both theoretical and experimental studies, which are also presented in Fig.~\ref{Fig:Vcd}. Those results originate from the CLEO'09~\cite{CLEO:2009svp}, BESIII'~\cite{BESIII:2013iro,BESIII:2015tql, BESIII:2017ylw, BESIII:2018xre, BESIII:2018eom, FlavourLatticeAveragingGroup:2019iem, Ablikim:2020hsc}, PDG'24~\cite{ParticleDataGroup:2024cfk}, FLAG'24~\cite{FlavourLatticeAveragingGroupFLAG:2024oxs}, LQCD~\cite{Ke:2023qzc, FermilabLattice:2022gku, Riggio:2017zwh, Lubicz:2017syv, Na:2011mc, FermilabLattice:2004ncd}, CLEO'09~\cite{CLEO:2009svp}, BaBar'14~\cite{BaBar:2014xzf} and FLAG'21 \cite{FlavourLatticeAveragingGroupFLAG:2021npn}. From the figure, our two methods yield results that demonstrate strong consistency with the majority of existing experimental measurements and theoretical predictions. The Scheme I result is in good agreement with FLAG'24, BESIII'17 and FLAG'21; the Scheme II result aligns closely with the  prediction from PDG'24, BESIII'15, BESIII(II)'18, LQCD'04, LQCD'11, LQCD'18, LQCD'19 and LQCDfit'23. Notably, the error bars of Scheme I and II exhibit substantial overlap, which validates the reliability of our computational framework. Furthermore, the results obtained by the two methods are highly consistent with those of PDG within the error range, which further proves the reliability of these two methods.

\section{Summary}\label{Sec:4}

In this paper, we investigated the semileptonic decay $D \to \pi$. Firstly, the TFF is calculated using the right-chiral and left-chiral currents correlators function within the framework of LCSR. For the main nonperturbative LCDAs in TFFs, we derive $\phi_{2;\pi}(x,\mu)$, $\phi^{p}_{3;\pi}(x,\mu)$, and $\phi^{\sigma}_{3;\pi}(x,\mu)$ by constructing the LCHO model, whose specific behaviors are presented in Fig.~\ref{Fig:DA-R} and Fig.~\ref{Fig:DA-L}, respectively. The relevant numerical results of $f_+^{D\pi}(0)$ are listed in Fig.~\ref{Fig:TFF0}. The results from the two methods are consistent with each other within errors and in good agreement with most existing experimental and theoretical results, confirming the reliability of our approach. Subsequently, the TFFs are extrapolated to the high $ q^2 $ region using $z(q^2,t)$ to converge the SSE. As illustrated in Fig.~\ref{Fig:TFF}, including results from other groups for comparison, our two curves are relatively close to the existing data within error bars.

After extrapolating the TFFs, the differential decay width of $D \to \pi$ is obtained and presented in Fig.~\ref{Fig:decaywidth}. Both curves from Schemes I and II are highly consistent with most experimental points. Then the corresponding branching fractions are also listed in Table~\ref{tab:fpi_values}. The results from both methods are consistent with the PDG values within errors and show good agreement with experimental measurements and LQCD theoretical predictions. Then, the branching fractions $\mathcal{B}(D^0 \to \pi^- e^+ \nu_e)$ from PDG'24~\cite{ParticleDataGroup:2024cfk} is used to calculate the CKM matrix $|V_{cd}|$. The results obtained from both methods align well with the results from LQCDfit'23, FLAG, BESIII, and the PDG'24.

Overall, in this paper, for the $D \to \pi$ decay process, despite differences in the results of the two methods, both agree well with experimental and theoretical results within uncertainties, thus verifying the feasibility of the two approaches. However, discrepancies still exist between some current experimental and theoretical results. With the continuous improvement of experimental measurement precision and theoretical calculation accuracy, we eagerly anticipate that this decay channel will be further explored by experimental collaborations, yielding more precise results in the near future. In addition, in the TFF calculation using the left-chiral correlator function, NLO corrections have not been incorporated due to the insufficient maturity of twist-3 LCDA calculations. Moving forward, with improvements in the calculation precision of twist-3 LCDAs, the refinement of their corresponding NLO corrections may help enhance the precision in determining LCDAs and related parameters, thereby further reducing theoretical uncertainties.
\\

\section{Acknowledgments}
This work was supported in part by the National Natural Science Foundation of China under Grant No.12265010, the Project of Guizhou Provincial Department of Science and Technology under Grants No.MS[2025]219 and No.CXTD[2025]030.

\end{document}